\documentstyle[12pt]{article}

\begin{document}

\rightline{FNT/T-96/10}
\rightline{BHAM-HEP/96-01}
\vspace{0.5cm}

\noindent
{\Large{\bf {\tt WWGENPV}~2.0 - A Monte Carlo Event Generator for
Four-Fermion Production at $e^+ e^-$ Colliders}}

\vspace{0.5cm}
\medskip
\noindent

\begin{center}
David G. CHARLTON$^{a}$, Guido MONTAGNA$^{b,c}$, \\
Oreste NICROSINI$^{c,b}$ and Fulvio PICCININI$^{c}$

\medskip
{\it $^a$Royal Society University Research Fellow, School of Physics
         and Space Research, University of Birmingham,
         Birmingham B15 2TT, UK} \\
{\it $^b$Dipartimento di Fisica Nucleare e Teorica, Universit\`a di
         Pavia, via A. Bassi n. 6 - 27100 PAVIA - ITALY} \\
{\it $^c$ INFN, Sezione di Pavia, via A. Bassi n. 6 -
27100 PAVIA - ITALY}
\end{center}

\bigskip
\noindent
Program classification: 11.1 \\
\bigskip

\noindent
{\small The Monte Carlo program {\tt WWGENPV}, designed for computing
distributions and generating events for
four-fermion production in $e^+ e^- $ collisions, is
described. The new version, 2.0, includes the full set of
the electroweak (EW) tree-level
matrix elements for double- and single-$W$ production, initial- and final-state
photonic radiation including $p_T / p_L$ effects in the Structure Function
formalism, all the relevant non-QED corrections (Coulomb correction, naive
QCD, leading EW corrections). An hadronisation interface to
{\tt JETSET} is also provided.
The program
 can be used in a three-fold way: as a Monte Carlo integrator for
weighted events, providing predictions for several observables relevant for
$W$ physics; as an adaptive integrator, giving  predictions for cross
sections, energy and invariant mass losses with high numerical precision;
as an event generator for
unweighted events, both at partonic and hadronic level. In all the branches,
the code can provide accurate and fast results. }


\vfil

\leftline{FNT/T-96/10}
\leftline{BHAM-HEP/96-01}
\eject

\leftline{\Large{\bf NEW VERSION SUMMARY}}
\vskip 15pt

\leftline{{\it Title of new version:} {\tt WWGENPV}~2.0}
\vskip 8pt

\leftline{{\it Catalogue number:} }
\vskip 8pt

\noindent
{\it Program obtainable from:} CPC Program Library, Queen's
University of Belfast, N. Ireland (see application form in this issue)
\vskip 8pt

\noindent
{\it Reference to original program:} {\tt WWGENPV}; {\it Cat.~no.:} ACNT;
{\it Ref. in CPC: } {\bf 90} (1995) 141
\vskip 8pt

\noindent
{\it Authors of original program:} Guido Montagna, Oreste Nicrosini
and Fulvio Piccinini
\vskip 8pt

\noindent
{\it The new version supersedes the original program}
\vskip 8pt

\noindent
\leftline{{\it Licensing provisions:} none}
\vskip 8pt

\noindent
{\it Computer for which the new version is designed:}
DEC ALPHA 3000, HP 9000/700 series;
{\it Installation:} INFN,
Sezione di Pavia, via A.~Bassi 6, 27100 Pavia, Italy
\vskip 8pt

\leftline{{\it Operating system under which the new version has been tested:}
VMS, UNIX}
\vskip 8pt

\leftline{{\it Programming language used in the new version:} FORTRAN 77}
\vskip 8pt

\noindent
\leftline{{\it Memory required to execute with typical data:} 
$\approx$~250 -- 450Kb }
\vskip 8pt

\leftline{{\it No. of bits in a word: } 32}
\vskip 8pt

\leftline{{\it No. of processors used: } 1}
\vskip 8pt

\leftline{\it The code has not been vectorized}
\vskip 8pt

\leftline{{\it Subprograms used:}
{\tt CERNLIB}~[1],  {\tt NAGLIB}~[2], {\tt JETSET}~[3],
{\tt RANLUX}~[4]  }
\vskip 8pt

\leftline{{\it No. of lines in distributed program, including test data,
etc.:} $\approx$~14000}

\vskip 8pt
\leftline{\it Correspondence to:}
\leftline{MONTAGNA@PAVIA.PV.INFN.IT;}
\leftline{NICROSINI@PAVIA.PV.INFN.IT; NICROSINI@VXCERN.CERN.CH; }
\leftline{PICCININI@PAVIA.PV.INFN.IT;}
\leftline{DAVE.CHARLTON@CERN.CH}
\vskip 8pt


\vskip 8pt

\leftline{{\it Nature of physical problem} }
\noindent
The precise measurement of the $W$-boson mass $M_W$ constitutes a
primary task
of the forthcoming experiments at the high energy electron--positron
collider LEP2 ($2 M_W \leq \sqrt{s} \leq 210$~GeV).
A meaningful comparison between theory and
experiment requires an accurate description of
the fully exclusive processes $e^+ e^-  \to 4f$, including the
main effects of radiative corrections, with the final goal of
providing predictions for the distributions
measured by the experiments.

\vskip 8pt

\leftline{{\it Method of solution} }
\noindent
Same as in the original program, as far as weighted event integration and
unweighted event generation are concerned. Adaptive Monte Carlo integration
for high numerical precision purposes is added. Optional hadronic interface
in the generation branch is supplied.

\vskip 8pt

\leftline{\it Reasons for the new version}
\noindent
The most promising methods for measuring the $W$-boson mass at LEP2 are the so
called ``threshold'' and ``direct reconstruction'' methods~[5].
For the first one,
a precise evaluation of the threshold cross section is required. For the
second one, a precise description of the invariant-mass shape of the hadronic
system in semileptonic decays is mandatory. In order to meet these
requirements, the previous version of the program has been improved by
extending the class of the tree-level EW diagrams taken into account, by
including $p_T / p_L $ effects both in initial- and final-state QED radiation,
by supplying an hadronic interface in the generation branch.

\vskip 8pt

\leftline{{\it Restrictions on the complexity of the problem} }
\noindent
While the semileptonic decay channels are complete at the level of the Born
approximation EW diagrams ({\tt CC11/CC20} diagrams), 
neutral current backgrounds are neglected in the
fully hadronic and leptonic decay channels. QED radiation is treated at the
leading logarithmic level. Due to the absence of a complete ${\cal O}
(\alpha)$/${\cal O} (\alpha_s)$ diagrammatic calculation, the most relevant EW
and QCD corrections are effectively incorporated according to the recipe given
in~[6]. No anomalous coupling effects are at present taken into account.

\vskip 8pt

\leftline{{\it Typical running time} }
\noindent
As adaptive integrator, the code provides cross section and energy and
invariant-mass losses with a relative accuracy of about 1\% in 8 min on 
HP 9000/735.
As integrator of weighted events, the code produces about 
$10^5$~events/min on the same system. The generation of a sample of $10^3$ 
hadronised unweighted events requires about 8 min on the same system.
\vskip 8pt

\vfil \eject
\leftline{{\it Unusual features of the program} }
\noindent
Subroutines from the library of mathematical subprograms {\tt NAGLIB}~[3]
for the numerical integrations
are used in the program, when the adaptive integration branch is selected.
\vskip 8pt

\leftline{{\it References} }
\noindent
[1] CERN Program Library, CN Division, CERN, Geneva.
\vskip 10pt\noindent
[2] NAG Fortran Library Manual Mark 16 (Numerical Algorithms
Group, Oxford, 1991).
\vskip 10pt\noindent
[3] T.~Sj\"ostrand, Comp. Phys. Commun. {\bf 82} (1994) 74; Lund University
Report LU TP 95-20 (1995).
\vskip 10pt\noindent
[4] F.~James, Comput. Phys. Commun. 79 (1994) 111.
\vskip 10pt\noindent
[5] {\it Physics at LEP2},  CERN Report 96-01, Theoretical Physics and
Particle Physics Experiments Divisions, G.~Altarelli, T.~Sj\"ostrand and
F.~Zwirner, eds., Vols. 1 and 2, Geneva, 19 February 1996.
\vskip 10pt\noindent
[6] W. Beenakker, F. Berends et al., ``WW cross-sections and distributions'',
in~[5],  Vol. 1, pag. 79.
\vskip 15pt
\vfil \eject

\leftline{\Large{\bf LONG WRITE-UP}}
\vskip 15pt

\noindent
\section{Introduction}
\vskip 10pt

The precise measurement of the $W$-boson mass $M_W$ constitutes a
primary task
of the forthcoming experiments at the high energy electron--positron
collider LEP2 ($2 M_W \leq \sqrt{s} \leq 210$~GeV).
A meaningful comparison between theory and
experiment requires an accurate description of
the fully exclusive processes $e^+ e^-  \to 4f$, including the
main effects of radiative corrections, with the final goal of
providing predictions for the  distributions
measured by the experiments. A large effort in the direction of developing
tools  dedicated to the investigation of
this item has been spent within the Workshop
``Physics at LEP2'', held at CERN during 1995. Such an effort has led to the
development of several independent four-fermion codes, both semianalytical and
Monte Carlo, extensively documented in~\cite{wweg}. {\tt WWGENPV} is one of
these codes, and the aim of the present paper is to describe in some detail
the developments performed with respect to the original version~\cite{cpcww},
where a description of the
formalism adopted and the physical ideas behind it can be found.

As discussed in~\cite{wmass}, the most promising methods for measuring the
$W$-boson mass at LEP2 are the so
called ``threshold'' and ``direct reconstruction'' methods.
For the first one,
a precise evaluation of the threshold cross section is required. For the
second one, a precise description of the invariant-mass shape of the hadronic
system in semileptonic and hadronic decays is mandatory. In order to meet these
requirements, the previous version of the program has been improved, both from
the technical and physical point of view.

On the technical side, in addition
to the ``weighted event integration''  and
``unweighted event generation'' branches,  the present version can
also be run as an  ``adaptive Monte Carlo'' integrator, in order to obtain
high numerical precision results for cross sections and other relevant
observables. In the ``weighted event integration'' branch, a ``canonical''
output can be selected, in which several observables are processed in parallel
together with their most relevant moments~\cite{wweg}.
Moreover, the program offers the possibility of generating events
according to a specific flavour quantum number assignment for the final-state
fermions, or of generating ``mixed samples'', namely a fully leptonic, fully
hadronic or semileptonic sample.

On the physical side, the class of tree-level EW diagrams taken into
account has been extended to include all the single resonant diagrams
({\tt CC11/CC20}), in such a way that all the 
charged current processes are covered.
Motivated by the physical relevance of keeping under control the effects of
the transverse degrees of freedom of photonic radiation, both for the $W$ mass
measurement and for the detection of anomalous couplings,
the contribution of QED radiation has been fully developed in the leading
logarithmic approximation, going beyond the initial-state, strictly collinear
approximation, to include $p_T / p_L $ effects both for initial- and
final-state photons. Last, an hadronic interface to {\tt JETSET}
in the generation branch has been added.

In the present version the neutral current backgrounds are neglected in the
fully hadronic and leptonic decay channels, but this is not a severe
limitation of
the program since, at least in the LEP2 energy range, these backgrounds
can be suppressed by means of proper invariant mass cuts. On the other hand,
the semileptonic decay channels are complete at the level of the Born
approximation EW diagrams ({\tt CC11/CC20} diagrams), 
and this feature allows to treat at best those
channels that are expected to be the most promising for the direct
reconstruction of the $W$ mass, because free of systematics such as the
``color reconnection'' and the ``Bose-Einstein correlation'' problems.

In the development of the code, particular attention has been paid to the
possibility of obtaining precise results in relatively short CPU time. As shown
in~\cite{wweg}, {\tt WWGENPV} is one of the most precise four fermion Monte
Carlo's from the numerical point of view. This feature allows the use of the
code also for fitting purposes.

\noindent
\section{The most important new features}
\vskip 10pt

In the following we list and briefly describe the most important
technical and physical developments implemented in the new version 
 of {\tt WWGENPV}.

\vskip 24pt \noindent
TECHNICAL IMPROVEMENTS

The present version of the program  consists of three branches, two
of them already present in the original version but upgraded in some
respect, the third one completely new.

\noindent

\begin{itemize}
\item Unweighted event generation branch.
This branch, meant for simulation purposes, has been
improved by supplying an option for an hadronisation interface
(see more details later on).

\item Weighted event integration branch.
This branch, intended for
computation only, includes as a new feature an option for
selecting a ``canonical'' output containing predictions for several
observables and their most relevant moments together with a
Monte Carlo estimate of the errors. According to the strategy
adopted in~\cite{wweg}, the first four Chebyshev/power
moments of the following quantities are computed:
the production angle of the $W^+$ with respect to the positron beam
({\tt TNCTHW, N=1,2,3,4}), the production angle $\vartheta_{d}$
of the down fermion with respect to the positron beam ({\tt TNCTHD}),
the decay angle $\vartheta_{d}^*$ of the down fermion with respect
to the direction of the decaying $W^-$ measured in its rest frame
({\tt TNCTHSR}), the energy $E_{d}$ of the down fermion 
 normalized to the beam energy ({\tt XDN}),
the sum of the energies of all radiated photons ({\tt XGN}) 
normalized to the beam energy,
the lost and visible photon energies normalized to the beam energy 
({\tt XGNL} and {\tt XGNV}), respectively, and, finally
\begin{eqnarray*}
<x_m> \, = \, {1 \over \sigma} \, \int \left( { { \sqrt{s_+} +
\sqrt{s_-} -2 M_W} \over {2 E_b} } \right) \, d \sigma
\end{eqnarray*}
where $s_+$ and $s_-$ are the invariant masses of the $W^+$ and $W^-$
decay products, respectively ({\tt XMN}).

\item Adaptive integration branch.
This new branch is intended for computation
only, but offers high precision performances. On top of the importance
sampling, an adaptive Monte Carlo integration algorithm is used. The
code returns the value of the cross section together with a Monte Carlo
estimate of the error. Moreover, if QED corrections are taken into 
account, also the
average energy and invariant mass losses are printed. The program must
be linked to NAG library for the Monte Carlo adaptive routine. Full
consistency between non-adaptive and adaptive integrations has been
explicitly proven.
\end{itemize}

In each of these three branches the user is asked to specify the
four-fermion final state which is required. The final states at present
available are those containing {\tt CC03} diagrams as a subset.
Their list, as appears
when running the code, is the following:

\begin{verbatim}
PURELY LEPTONIC PROCESSES
[0]  ---> E+ NU_E E- BAR NU_E
[1]  ---> E+ NU_E MU- BAR NU_MU
[2]  ---> E- BAR NU_E MU+ NU_MU
[3]  ---> E+ NU_E TAU- BAR NU_TAU
[4]  ---> E- BAR NU_E TAU+ NU_TAU
[5]  ---> MU+ NU_MU MU- BAR NU_MU
[6]  ---> MU+ NU_MU TAU- BAR NU_TAU
[7]  ---> MU- BAR NU_MU TAU+ NU_TAU
[8]  ---> TAU+ NU_TAU TAU- BAR NU_TAU
SEMILEPTONIC PROCESSES
[9]  ---> E+ NU_E D BAR U
[10] ---> E- BAR NU_E BAR D U
[11] ---> E+ NU_E S BAR C
[12] ---> E- BAR NU_E BAR S C
[13] ---> MU+ NU_MU D BAR U
[14] ---> MU- BAR NU_MU BAR D U
[15] ---> MU+ NU_MU S BAR C
[16] ---> MU- BAR NU_MU BAR S C
[17] ---> TAU+ NU_TAU D BAR U
[18] ---> TAU- BAR NU_TAU BAR D U
[19] ---> TAU+ NU_TAU S BAR C
[20] ---> TAU- BAR NU_TAU BAR S C
HADRONIC PROCESSES
[21] ---> D BAR U BAR D U
[22] ---> D BAR U BAR S C
[23] ---> S BAR C BAR D U
[24] ---> S BAR C BAR S C
MIXED SAMPLES
[25] ---> LEPTONIC SAMPLE
[26] ---> SEMILEPTONIC SAMPLE
[27] ---> HADRONIC SAMPLE
\end{verbatim}
It is worth noting that in the weighted event integration and unweighted
event
generation branches, besides the possibility of selecting a specific
four-fermion final state, an option is present for considering  three
realistic mixed samples corresponding to
the fully leptonic, fully hadronic or semileptonic decay channel,
respectively.
When the generation of a mixed sample is required,
as a a first step the cross section for
each contributing channel is calculated; as a second step, the unweighted
events are generated for each contributing channel with a frequency
given by the weight of that particular channel with respect to the total.

In the generation branch, if the hadronisation interface is not enabled,
an $n$-tuple is created with the following structure:

\begin{verbatim}
'X_1','X_2','EB'         ! x_{1,2} represent the
                         ! energy fractions of incoming e^- and
                         ! e^+ after ISR; EB is the beam energy;
'Q1X','Q1Y','Q1Z','Q1LB' ! x,y,z components of the momentum
                         ! of particle 1 and the particle label
                         ! according to PDG; the final-state
                         ! fermions are assumed to be massless;
'Q2X','Q2Y','Q2Z','Q2LB' ! as above, particle 2
'Q3X','Q3Y','Q3Z','Q3LB' ! "  "          "    3
'Q4X','Q4Y','Q4Z','Q4LB' ! "  "          "    4
'AK1X','AK1Y','AK1Z'     ! x,y,z components of the momentum of
                         ! the photon from particle 1;
                         ! they are 0 if no FSR has been chosen
                         ! and/or if particle 1 is a neutrino;
'AK2X','AK2Y','AK2Z'     ! as above, particle 2
'AK3X','AK3Y','AK3Z'     ! "  "          "    3
'AK4X','AK4Y','AK4Z'     ! "  "          "    4
'AKEX','AKEY','AKEZ'     ! x,y,z components of the momentum of
                         ! the photon from the initial-state
                         ! electron; they are 0 if no ISR has
                         ! been chosen;
'AKPX','AKPY','AKPZ'     ! as above, initial-state positron;
\end{verbatim}

If the hadronisation interface is enabled, fully hadronised events are
instead made available to a user routine in the {\tt /HEPEVT/} format
(see below).

\vskip 24pt \noindent
PHYSICAL IMPROVEMENTS \\

The main theoretical developments with respect to the original version
concern the inclusion of additional matrix elements to
the tree-level kernel and  a more sophisticated treatment of the photonic
radiation, beyond the initial-state, strictly collinear approximation.
Moreover, an hadronisation interface to { \tt JETSET} has been also provided. 

{\it Tree-level EW four-fermion diagrams} -- In addition to double-resonant
charged-current diagrams {\tt CC03}
already present in the previous version,
the matrix element includes also the
single-resonant charged-current diagrams {\tt CC11} for $\mu $ and $\tau$'s in
the final state, 
and {\tt CC20} for final states containing electrons.
This allows a complete treatment at the level of four-fermion
EW diagrams of the semileptonic sample, which appears the most
promising and cleanest for the direct mass reconstruction
method due to the absence of potentially large ``interconnection''
effects~\cite{wmass}.
Concerning {\tt CC20} diagrams,
the importance sampling technique has been extended to
take care of the peaking behaviour of the matrix element
when small momenta of the virtual photon are involved. As a consequence
of the fact that the tree-level matrix-element is
computed in the massless limit, a cut on the minimum electron (positron)
scattering angle must be imposed. The inclusion of such a cut
eliminates the problems connected with gauge-invariance in the case of
{\tt CC20} processes, for which the present version does not
include any so-called ``reparation'' scheme~\cite{bhf,wwcd}. Anyway, when
for instance a set of ``canonical cuts''~\cite{wweg} is used,
the numerical relevance of
such gauge-invariance restoring schemes has been shown~\cite{bhf,wwcd} to be
negligible compared with the expected experimental accuracy.

{\it Photonic corrections} -- As far as photonic effects are concerned, 
the original version, as
stated above, included only leading logarithmic initial-state corrections
in the collinear approximation within the SF formalism. The
treatment has been extended in a two-fold way: the contribution of
final-state radiation has been included and the $p_T / p_L$ effects
have been implemented both for initial- and final-state radiation. The
inclusion of the transverse degrees of freedom has been achieved by generating
the fractional energy $x_{\gamma}$ of the radiated photons by means
of resummed electron structure
functions $D(x; s)$~\cite{sf} ($x_{\gamma} = 1 - x$) and the angles
using an angular factor inspired by the pole behaviour
$1 / (p \cdot k)$ for each charged emitting fermion. This allows
to incorporate leading QED radiative corrections originating from
infrared and collinear singularities, taking into account at the same
time the dominant kinematic effects due to non-strictly collinear
photon emission, in such a way that the universal factorized photonic spectrum
is recovered.  According to this procedure, the
leading logarithmic corrections from initial- and final-state radiation
are isolated as a gauge-invariant
subset of the full calculation (not yet available) of the
electromagnetic corrections to $e^+ e^- \to 4f$.
Due to the inclusion of $p_T$-carrying photons at
the level of initial-state radiation, the Lorentz boost allowing the
reconstruction of the hard-scattering event from the c.m. system to
the laboratory one has been generalized to keep under control the
$p_T$ effects on the beam particles.
Final state radiation and $p_T / p_L$ effects are not taken into account
in the adaptive integration branch.

{\it Non-QED corrections} -- Coulomb correction is treated as in
the original version
on double-resonant {\tt CC03} diagrams. QCD corrections are
implemented in the present version in the  naive form
according to the recipe described in~\cite{wweg,wwcd}. The
treatment of the leading EW contributions is unchanged with respect to the
original version.

{\it Hadronisation} -- Final-state quarks
issuing from the electroweak 4-fermion scattering are
not experimentally observable.
An hadronisation interface is provided to the {\tt JETSET} package~\cite{jetset}
to allow events to be extrapolated to the hadron level, for example for
input to a detector simulation program.
Specifically, the 4-fermion event structure is converted to the
{\tt /HEPEVT/} convention, then {\tt JETSET} is called to simulate QCD
partonic evolution (via routines {\tt LUJOIN} and {\tt LUSHOW}) and
hadronisation (routine {\tt LUEXEC}).
In making this conversion, masses must be added to the outgoing fermions,
considered massless in the hard scattering process.
This is done by rescaling the fermion momenta by a single scale factor,
keeping the flight directions fixed in the rest frame of the four fermion
system.
In the QCD evolution phase, strings join quarks coming from the same W decay. 
The virtuality scale of the QCD evolution is taken to be the invariant
mass-squared of each evolving fermion pair.
No colour reconnection is included by default, although it could be implemented
by appropriate modification of routine {\tt WWGJIF} if required. 
Bose - Einstein correlations are neglected in the present version.
The resulting event structure is then made available to the user
in the {\tt /HEPEVT/} common block via a routine
{\tt WWUSER} for further analysis, such as writing out for later
input to a detector simulation program.
The {\tt WWUSER} routine is also called at program initialisation time to
allow the user to set any non-standard {\tt JETSET} program options, for
example, and at termination time to allow any necessary clean-up.
A dummy {\tt WWUSER} routine is supplied with the program.
The only {\tt JETSET} option which is changed by 
{\tt WWGENPV} from its default value
controls emission of gluons and photons by final-state
partons\footnote{\footnotesize {\tt JETSET} parameter {\tt MSTJ(41)}},
turning off final-state photon emission
simulation from {\tt JETSET} if activated in {\tt WWGENPV}, 
to avoid double counting.
\\

All the new features of the program can be switched on/off by means of
separate flags, as described in the following.

\section{Input}

\vskip 10pt

Here we give a short explanation of the input parameters and flags
required when running the program.

\noindent
{\bf \begin{verbatim}
OGEN(CHARACTER*1)
\end{verbatim}}
\noindent
It controls the use of the program as a Monte Carlo
event generator of unweighted events ({\tt OGEN = G})
or as a Monte Carlo/adaptive integrator
for weighted events ({\tt OGEN = I}).
\vskip 5pt

\noindent
{\bf \begin{verbatim}
RS(REAL*8)
\end{verbatim}}
\noindent
The centre-of-mass energy (in GeV).

\vskip 5pt

\noindent
{\bf \begin{verbatim}
OFAST(CHARACTER*1)
\end{verbatim}}
\noindent
It selects ({\tt OFAST = Y}) the adaptive integration branch, when
{\tt OGEN = I}. When this choice is done, the required relative accuracy
 of the numerical integration has to be supplied by means of the {\tt REAL*8}
 variable {\tt EPS}.

\noindent
{\bf \begin{verbatim}
NHITWMAX(INTEGER)
\end{verbatim}}
\noindent
Required by the Monte Carlo integration branch. It is the maximum number of
calls for the Monte Carlo loop.
\vskip 5pt

\noindent
{\bf \begin{verbatim}
NHITMAX(INTEGER)
\end{verbatim}}
\noindent
Required by the event-generation branch. It is the maximum number of
hits for the hit-or-miss procedure.
\vskip 5pt

\noindent
{\bf \begin{verbatim}
IQED(INTEGER)
\end{verbatim}}
\noindent
This flag allows the user to switch on/off the contribution of the
initial-state radiation. If {\tt IQED = 0} the distributions are
computed in lowest-order approximation, while for {\tt IQED = 1}
the QED corrections are included in the calculation.

\vskip 5pt

\noindent
{\bf \begin{verbatim}
OPT(CHARACTER*1)
\end{verbatim}}
\noindent
This flag controls the inclusion of $p_T / p_L$ effects for
the initial-state radiation. It is ignored in the adaptive integration
branch where only initial-state strictly collinear radiation is allowed.

\vskip 5pt

\noindent
{\bf \begin{verbatim}
OFS(CHARACTER*1)
\end{verbatim}}
\noindent
It is the option for including final-state radiation. It is assumed 
that final-state radiation can be switched on only if initial-state 
radiation including $p_T / p_L$ effects is on, in which case 
final-state radiation includes $p_T / p_L$ effects as well.
Ignored in the adaptive integration branch.

\vskip 5pt

\noindent
{\bf \begin{verbatim}
ODIS(CHARACTER*1)
\end{verbatim}}
\noindent
Required by the integration branch.
It selects the kind of experimental distribution. For {\tt ODIS = T}
the program computes the total cross section
(in pb) of the process;
 for {\tt ODIS = W} the value of the invariant-mass distribution 
 $d \sigma / d M$ of the system $d \bar u$ ({\tt IWCH = 1}) or of the system
 ${\bar d} u$ ({\tt IWCH = 2}) is returned (in pb/GeV).

\vskip 5pt

\noindent
{\bf \begin{verbatim}
OWIDTH(CHARACTER*1)
\end{verbatim}}
\noindent
It allows a different choice of the value of the $W$-width.
{\tt OWIDTH = Y} means that the tree-level Standard Model
formula for the $W$-width is used; {\tt OWIDTH = N} requires that
the $W$-width is supplied by the user in GeV.
\vskip 15pt

\vfil\eject
\noindent
{\bf \begin{verbatim}
NSCH(INTEGER)
\end{verbatim}}
\noindent
The value of
{\tt NSCH} allows the user to choose the calculational scheme for the
weak mixing angle and the gauge coupling. Three choices are available. If
{\tt NSCH=1}, the input parameters
used are $G_F, M_W, M_Z$ and the calculation is
performed at tree level. If {\tt NSCH = 2} or
{\tt 3}, the input parameters used are
$\alpha(Q^2), G_F, M_W$ or $\alpha(Q^2), G_F, M_Z$, respectively,
 and the calculation is performed using the QED coupling constant at a
proper scale $Q^2$, which is requested as further input. The 
recommended choice is {\tt NSCH = 2}, consistently with~\cite{wweg}.

\noindent
{\bf \begin{verbatim}
OCOUL(CHARACTER*1)
\end{verbatim}}
\noindent
This flag allows the user to switch on/off the contribution of the
Coulomb correction. Unchanged with respect to the old version of 
the program.

\noindent
{\bf \begin{verbatim}
OQCD(CHARACTER*1)
\end{verbatim}}
\noindent
This flag allows the user to switch on/off the contribution of the naive
QCD correction.

\noindent
{\bf \begin{verbatim}
ICHANNEL(INTEGER)
\end{verbatim}}
\noindent
A channel corresponding to a specific flavour quantum number assignment
can be chosen.

\noindent
{\bf \begin{verbatim}
ANGLMIN(REAL*8)
\end{verbatim}}
\noindent
The minimum electron (positron) scattering angle (deg.) in the laboratory
frame. It is ignored when {\tt CC20} graphs are not selected.

\noindent
{\bf \begin{verbatim}
SRES(CHARACTER*1)
\end{verbatim}}
\noindent
Option for switching on/off single-resonant diagrams ({\tt CC11}).

\noindent
{\bf \begin{verbatim}
OCC20(CHARACTER*1)
\end{verbatim}}
\noindent
Option for switching on/off single-resonant diagrams when
electrons (positrons) occur in the final state ({\tt CC20}).

\noindent
{\bf \begin{verbatim}
OOUT(CHARACTER*1)
\end{verbatim}}
\noindent
Option for ``canonical'' output containing results for several observables
and their most important moments. It is active only in the Monte Carlo
integration branch.

\noindent
{\bf \begin{verbatim}
OHAD(CHARACTER*1)
\end{verbatim}}
\noindent
Option for switching on/off hadronisation interface in the
unweighted event generation branch.

\section{Test run output}

\vskip 10pt

The typical new calculations that can be performed with the 
updated version of the
program are illustrated in the following examples.

\vskip 8pt\noindent
\leftline{\bf Sample 1} An example of adaptive integration is 
provided. The process considered is $e^+ e^- \to e^+ \nu_e d \bar u$ 
({\tt CC20}). 
The output gives the cross section, together with the energy and 
invariant-mass losses from initial-state radiation. ``Canonical'' cuts 
are imposed as in~\cite{wweg}. The input card is as follows: 
\begin{verbatim}
OGEN = I
RS = 190.D0
OFAST = Y
EPS = 1.D-2
IQED = 1
OPT = N
OFS = N
ODIS = T
OWIDTH = Y
NSCH = 2
ALPHM1 = 128.07D0
OCOUL = N
OQCD = N
ICHANNEL = 9
ANGLMIN = 10.D0
SRES = Y
OCC20 = Y
OOUT = N
\end{verbatim}

\vskip 8pt\noindent
\leftline{\bf Sample 2} An example of weighted event integration 
is provided. Here the process considered is $e^+ e^- \to \mu^+ \nu_{\mu} 
d \bar u$ ({\tt CC11}). ``Canonical'' cuts 
are imposed as before. The ``canonical'' output is provided. The input card
differs from the previous one as follows: 
\begin{verbatim}
RS = 175.D0
OFAST = N
NHITWMAX = 100000
OPT = Y 
OFS = Y
OCOUL = Y
OQCD = Y
ICHANNEL = 13
OCC20 = N
OOUT = Y
\end{verbatim}

\vskip 8pt\noindent
\leftline{\bf Sample 3} An example of unweighted event generation including 
hadronisation is provided. A sample of 100 events 
corresponding to the full semileptonic channel is generated. The detailed 
list of an hadronised event is given. The input card
differs from the first one as follows: 
\begin{verbatim}
OGEN = G
RS = 175.D0
NHITMAX = 100
OPT = Y 
OFS = Y
OCOUL = Y
OQCD = Y
ICHANNEL = 26
ANGLMIN = 5.D0
OHAD = Y
\end{verbatim}
\vskip 10pt

\vskip 15pt

\section{Conclusions}

The program {\tt WWGENPV 2.0} has been described. In its present version it
allows the treatment of all the four-fermion reactions including the {\tt CC03}
class of diagrams as a subset. This means that all the semileptonic  channels
are complete from the tree-level diagrams point of view, whereas fully leptonic
and fully hadronic channels are treated in the CC approximation. Since the most
promising channels for the $W$ mass reconstruction are the semileptonic ones,
the present version of the code allows a precise analysis of such data.
Moreover, NC backgrounds can be suppressed by proper invariant mass cuts, so
that the code is also usable for fully hadronic and leptonic events analysis,
with no substantial loss of reliability. Initial- and final-state QED radiation
is taken into account within the SF formalism, including finite $p_T / p_L$
effects in the leading logarithmic approximation. The Coulomb correction is
included for the {\tt CC03 } graphs. Naive QCD and leading EW corrections
are implemented as well. An hadronic interface to 
{\tt JETSET} is also provided.

The code as it stands is a valuable tool for the analysis of LEP2 data, with
particular emphasis to the threshold and direct reconstruction methods for the
measurement of the $W$-boson mass. Speed and high numerical accuracy allow the
use of the program also for fitting purposes.

The code is supported. Future releases of {\tt WWGENPV} will include:
\begin{itemize}

\item an interface to the code {\tt HIGGSPV}~\cite{wweg,egdp} in
order to treat all the possible
four-fermion processes in the massless limit, including Higgs-boson signals;

\item implementation of anomalous couplings;

\item implementation of CKM effects;

\item the extension of the hadronic interface to {\tt HERWIG}~\cite{herwig}.

\end{itemize}

\vfill\eject

\pagestyle{empty}
\noindent
\leftline{\Large \bf Test Run Output}
\vskip 10pt

\leftline{\bf Test run output 1}
{\footnotesize
\begin{verbatim}

 SQRT(S) =  190.0  GEV
 M_Z =  91.1888  GEV
 G_Z =  2.4974  GEV
 M_W =  80.23  GEV
 G_W =  2.03367033063195  GEV
 STH2 =  .2310309124510679
 GVE =  -1.409737267299689E-02
 GAE =  -.185794027211796
 GWF =  .230409927395451
 GWWZ =  -.571479454308384
  
   
 IFAIL=  0
 NCALLS =  120592
 XSECT FOR WEIGHTED EVENTS
 XSECT =  .6133752577743349 +-  3.741996226354238E-03 (PB)
 
 IFAIL=  0
 NCALLS =  239448
 ENERGY LOSS
 ENL =  2.10061682664439 +-  3.009086376029191E-02 (GEV)
 
 IFAIL=  0
 NCALLS =  246756
 INVARIANT MASS LOSS
 IML =  2.08936495443206 +-  2.905625928853976E-02 (GEV)

\end{verbatim} }

\vfil\eject

\leftline{\bf Test run output 2}
{\footnotesize
\begin{verbatim}

 SQRT(S) =  175.0  GEV
 M_Z =  91.1888  GEV
 G_Z =  2.4974  GEV
 M_W =  80.23  GEV
 G_W =  2.03367033063195  GEV
 STH2 =  .2310309124510679

 GVE =  -1.409737267299689E-02
 GAE =  -.185794027211796
 GWF =  .230409927395451
 GWWZ =  -.571479454308384

 NCALLS =  95214 100000

 XSECT FOR WEIGHTED EVENTS

 EFF(Y) =  .9240040758891746
 EFF(N) =  .9704498034839147
 XSECT(Y) =  .4757101931318581 +-  2.533828802525367E-03 (PB)
 XSECT(N) =  .4839578519633417 +-  2.552018837243229E-03 (PB)

                       CANONICAL OUTPUT

 OBSERVABLE CUTS DIAGRAMS   RS      VALUE     MC ERROR

 SIGMA (PB)  Y     CC11    175.0  .4757101  2.53383E-03
 SIGMA (PB)  N     CC11    175.0  .4839578  2.55202E-03
 T1CTHW      Y     CC11    175.0  .3077387  5.94105E-03
 T2CTHW      Y     CC11    175.0  -.217813  2.65517E-03
 T3CTHW      Y     CC11    175.0  -.149287  4.88436E-03
 T4CTHW      Y     CC11    175.0  -.130416  3.47127E-03
 T1CTHW      N     CC11    175.0  .3102274  5.91031E-03
 T2CTHW      N     CC11    175.0  -.215368  2.65276E-03
 T3CTHW      N     CC11    175.0  -.149196  4.84336E-03
 T4CTHW      N     CC11    175.0  -.130814  3.43812E-03
 T1CTHMU     Y     CC11    175.0  .333774  5.53691E-03
 T2CTHMU     Y     CC11    175.0  -.318545  2.08559E-03
 T3CTHMU     Y     CC11    175.0  -.218236  5.39418E-03
 T4CTHMU     Y     CC11    175.0  -8.06308E-02  3.77469E-03
 T1CTHMU     N     CC11    175.0  .3325118  5.47350E-03
 T2CTHMU     N     CC11    175.0  -.32043  2.05915E-03
 T3CTHMU     N     CC11    175.0  -.21926  5.36138E-03
 T4CTHMU     N     CC11    175.0  -8.09428E-02  3.74979E-03
 T1CTHSR     Y     CC11    175.0  .148885  4.59988E-03
 T2CTHSR     Y     CC11    175.0  -.311119  2.31110E-03
 T3CTHSR     Y     CC11    175.0  -8.34748E-02  4.69059E-03
 T4CTHSR     Y     CC11    175.0  -7.81930E-02  3.84710E-03
 T1CTHSR     N     CC11    175.0  .1486743  4.56510E-03
 T2CTHSR     N     CC11    175.0  -.309691  2.30747E-03
 T3CTHSR     N     CC11    175.0  -8.26744E-02  4.65586E-03
 T4CTHSR     N     CC11    175.0  -7.80504E-02  3.81642E-03
 XMU1        Y     CC11    175.0  .5250259  5.99384E-03
 XMU2        Y     CC11    175.0  .2888953  3.56491E-03
 XMU3        Y     CC11    175.0  .1652849  2.21269E-03
 XMU4        Y     CC11    175.0  9.76726E-02  1.42417E-03
 XMU1        N     CC11    175.0  .525016  5.93588E-03
 XMU2        N     CC11    175.0  .2889403  3.53242E-03
 XMU3        N     CC11    175.0  .1653616  2.19389E-03
 XMU4        N     CC11    175.0  9.77576E-02  1.41285E-03
 XG1         Y     CC11    175.0  1.28590E-02  2.11471E-04
 XG2         Y     CC11    175.0  1.10076E-03  2.30737E-05
 XG3         Y     CC11    175.0  1.53589E-04  4.44690E-06
 XG4         Y     CC11    175.0  3.17221E-05  1.49813E-06
 XG1         N     CC11    175.0  1.28810E-02  2.09985E-04
 XG2         N     CC11    175.0  1.10500E-03  2.30841E-05
 XG3         N     CC11    175.0  1.55089E-04  4.49931E-06
 XG4         N     CC11    175.0  3.23389E-05  1.50661E-06
 XG1L        Y     CC11    175.0  1.28590E-02  2.11471E-04
 XG2L        Y     CC11    175.0  1.10076E-03  2.30737E-05
 XG3L        Y     CC11    175.0  1.53589E-04  4.44690E-06
 XG4L        Y     CC11    175.0  3.17221E-05  1.49813E-06
 XG1L        N     CC11    175.0  1.28810E-02  2.09985E-04
 XG2L        N     CC11    175.0  1.10500E-03  2.30841E-05
 XG3L        N     CC11    175.0  1.55089E-04  4.49931E-06
 XG4L        N     CC11    175.0  3.23389E-05  1.50661E-06
 XG1V        Y     CC11    175.0  .0  .0
 XG2V        Y     CC11    175.0  .0  .0
 XG3V        Y     CC11    175.0  .0  .0
 XG4V        Y     CC11    175.0  .0  .0
 XG1V        N     CC11    175.0  .0  .0
 XG2V        N     CC11    175.0  .0  .0
 XG3V        N     CC11    175.0  .0  .0
 XG4V        N     CC11    175.0  .0  .0
 XM1         Y     CC11    175.0  -1.26840E-02  3.40080E-04
 XM2         Y     CC11    175.0  1.99478E-03  5.72397E-05
 XM3         Y     CC11    175.0  -3.12925E-04  1.31033E-05
 XM4         Y     CC11    175.0  7.41696E-05  3.73455E-06
 XM1         N     CC11    175.0  -1.28271E-02  3.38555E-04
 XM2         N     CC11    175.0  2.02728E-03  5.73580E-05
 XM3         N     CC11    175.0  -3.24971E-04  1.33342E-05
 XM4         N     CC11    175.0  7.87204E-05  3.86960E-06
\end{verbatim} }

\vfil\eject

\leftline{\bf Test run output 3}

{\footnotesize
\begin{verbatim}

 HADRONISATION VIA JETSET
 WWGJIF: Initialising WWGENPV to JETSET interface
 PHOTON FSR WILL BE HANDLED BY WWGENPV:
 TURNING OFF PHOTON FSR FROM JETSET
1
 SQRT(S) =  175.0  GEV
 M_Z =  91.1888  GEV
 G_Z =  2.4974  GEV
 M_W =  80.23  GEV
 G_W =  2.08823657515165  GEV
 STH2 =  .2310309124510679

 GVE =  -1.409737267299689E-02
 GAE =  -.185794027211796
 GWF =  .230409927395451
 GWWZ =  -.571479454308384


                            Event listing (summary)

    I  particle/jet KS     KF orig    p_x      p_y      p_z       E        m

    1  !e+!         21    -11    0     .000     .000  -87.500   87.500     .001
    2  !e-!         21     11    0     .000     .000   87.500   87.500     .001
    3  !e+!         21    -11    1     .000     .000  -87.500   87.500     .001
    4  !e-!         21     11    2     .000     .000   87.500   87.500     .001
    5  gamma         1     22    1     .000     .000     .000     .000     .000
    6  nu_e          1     12    0   45.589   -8.504   -7.296   46.945     .000
    7  e+            1    -11    0  -32.518    1.009  -24.213   40.555     .001
    8  gamma         1     22    7    -.063    -.003    -.054     .083     .000
    9  (c~)      A  11     -4    0   -6.482    5.087  -14.629   16.844    1.350
   10  (g)       V  11     21    0    -.912    -.166   -1.330    1.621     .000
   11  (g)       A  11     21    0   -1.717    2.148     .149    2.754     .000
   12  (g)       V  11     21    0    -.404    -.150    -.153     .457     .000
   13  (g)       A  11     21    0     .524     .257    -.485     .759     .000
   14  (g)       V  11     21    0     .568    5.813   -3.909    7.028     .000
   15  (g)       A  11     21    0    -.833     .123    -.634    1.054     .000
   16  (g)       V  11     21    0   -1.616    -.218     .984    1.904     .000
   17  (g)       A  11     21    0    -.893    -.594    -.345    1.127     .000
   18  (g)       V  11     21    0    -.064   -1.133     .124    1.141     .000
   19  (u)       A  11      2    0   -1.275    -.657    4.823    5.032     .006
   20  (u~)      V  11     -2    0   -1.276     .758    3.201    3.528     .006
   21  (g)       A  11     21    0     .733    -.103    2.158    2.281     .000
   22  (g)       V  11     21    0     .691    -.234    2.326    2.438     .000
   23  (g)       A  11     21    0     .410   -1.788   16.927   17.027     .000
   24  (s)       V  11      3    0    -.461   -1.646   22.354   22.420     .199
   25  (string)     11     92    9  -13.104   10.511  -15.404   39.722   32.532
   26  (D~0)        11   -421   25   -6.057    4.073  -13.996   15.894    1.865
   27  (pi0)        11    111   25   -1.112    1.280    -.302    1.728     .135
   28  (rho-)       11   -213   25   -1.158    1.158   -2.172    2.803     .676
   29  (omega)      11    223   25    -.792     .558     .401    1.303     .774
   30  (Sigma*~+)   11  -3114   25     .123    2.038   -1.332    2.787    1.351
   31  (K*-)        11   -323   25     .319    1.963   -1.639    2.729     .896
   32  (Delta+)     11   2214   25    -.155    1.330   -1.314    2.263    1.266
   33  (rho-)       11   -213   25   -1.026     .221     .345    1.292     .670
   34  (pi0)        11    111   25    -.256     .036    -.546     .619     .135
   35  (rho0)       11    113   25    -.488    -.118     .538    1.103     .822
   36  pi+           1    211   25    -.420     .013    -.275     .521     .140
   37  pi-           1   -211   25    -.612    -.620     .002     .882     .140
   38  (rho+)       11    213   25    -.474   -1.082    1.048    1.662     .520
   39  (rho0)       11    113   25    -.997    -.339    3.839    4.136    1.122
   40  (string)     11     92   20     .096   -3.014   46.967   47.695    7.733
   41  p~-           1  -2212   40    -.233     .669    3.532    3.723     .938
   42  p+            1   2212   40     .259    -.320    2.633    2.826     .938
   43  (Delta~--)   11  -2224   40    -.208    -.282    7.298    7.431    1.357
   44  pi+           1    211   40     .568    -.518    3.830    3.909     .140
   45  (Lambda0)    11   3122   40    -.290   -2.562   29.673   29.806    1.116
   46  (K*+)        11    323   26   -3.673    2.525   -9.054   10.129     .875
   47  (rho-)       11   -213   26   -2.384    1.548   -4.942    5.765     .856
   48  gamma         1     22   27   -1.055    1.186    -.310    1.617     .000
   49  gamma         1     22   27    -.058     .094     .008     .111     .000
   50  pi-           1   -211   28    -.837    1.095   -1.923    2.370     .140
   51  (pi0)        11    111   28    -.321     .063    -.249     .433     .135
   52  pi+           1    211   29    -.610     .314     .197     .727     .140
   53  pi-           1   -211   29    -.236     .248     .217     .429     .140
   54  (pi0)        11    111   29     .055    -.005    -.014     .146     .135
   55  (Lambda~0)   11  -3122   30     .085    1.705    -.918    2.237    1.116
   56  pi+           1    211   30     .038     .333    -.414     .551     .140
   57  (K~0)        11   -311   31     .462    1.760   -1.507    2.414     .498
   58  pi-           1   -211   31    -.143     .203    -.133     .315     .140
   59  n0            1   2112   32     .123     .910   -1.024    1.665     .940
   60  pi+           1    211   32    -.277     .420    -.290     .598     .140
   61  pi-           1   -211   33     .005     .082     .159     .227     .140
   62  (pi0)        11    111   33   -1.031     .138     .186    1.065     .135
   63  gamma         1     22   34    -.201    -.017    -.301     .362     .000
   64  gamma         1     22   34    -.055     .053    -.245     .257     .000
   65  pi-           1   -211   35    -.071    -.134     .622     .655     .140
   66  pi+           1    211   35    -.417     .016    -.084     .448     .140
   67  pi+           1    211   38     .032    -.277     .386     .497     .140
   68  (pi0)        11    111   38    -.506    -.804     .661    1.166     .135
   69  pi-           1   -211   39    -.186     .113     .094     .275     .140
   70  pi+           1    211   39    -.811    -.452    3.745    3.861     .140
   71  p~-           1  -2212   43    -.135    -.544    5.493    5.600     .938
   72  pi-           1   -211   43    -.073     .262    1.805    1.831     .140
   73  n0            1   2112   45    -.262   -2.385   26.596   26.721     .940
   74  (pi0)        11    111   45    -.028    -.177    3.077    3.085     .135
   75  (K0)         11    311   46   -2.419    1.634   -6.536    7.176     .498
   76  pi+           1    211   46   -1.253     .891   -2.517    2.953     .140
   77  pi-           1   -211   47    -.404     .604   -1.633    1.792     .140
   78  (pi0)        11    111   47   -1.980     .945   -3.310    3.973     .135
   79  gamma         1     22   51    -.223     .051    -.091     .246     .000
   80  gamma         1     22   51    -.098     .012    -.158     .187     .000
   81  gamma         1     22   54     .048    -.066     .004     .082     .000
   82  gamma         1     22   54     .006     .062    -.017     .064     .000
   83  p~-           1  -2212   55     .064    1.274    -.756    1.755     .938
   84  pi+           1    211   55     .021     .431    -.161     .482     .140
   85  (K_S0)       11    310   57     .462    1.760   -1.507    2.414     .498
   86  gamma         1     22   62    -.111     .044    -.011     .120     .000
   87  gamma         1     22   62    -.920     .094     .197     .945     .000
   88  gamma         1     22   68    -.034    -.065     .012     .074     .000
   89  gamma         1     22   68    -.472    -.740     .649    1.091     .000
   90  gamma         1     22   74    -.035    -.088    2.449    2.451     .000
   91  gamma         1     22   74     .007    -.089     .628     .634     .000
   92  (K_S0)       11    310   75   -2.419    1.634   -6.536    7.176     .498
   93  gamma         1     22   78   -1.877     .868   -3.149    3.768     .000
   94  gamma         1     22   78    -.103     .077    -.160     .205     .000
   95  pi-           1   -211   85     .270     .383    -.379     .618     .140
   96  pi+           1    211   85     .192    1.377   -1.128    1.796     .140
   97  pi+           1    211   92    -.563     .463   -2.034    2.165     .140
   98  pi-           1   -211   92   -1.857    1.171   -4.502    5.011     .140
                   sum:   .00          .000     .000     .000  175.000  175.000
 NHIT =  100

 XSECT FOR UNWEIGHTED EVENTS

 EFF =  1.887112905965163E-03
 XSECT =  6.79360646147459 +-  .6787193283232224 (PB)
 NBIAS/NMAX =  .0
  USER TERMINATION
  A TOTAL OF  100 EVENTS WERE GENERATED
  THE TOTAL CROSS-SECTION WAS: 6.793606281280517E-09 MB
\end{verbatim}}


\begin{thebibliography}{99}

\bibitem{wweg} D. Bardin, R. Kleiss et al., ``Event generators for WW
physics'', in {\it Physics at LEP2}, CERN {\bf 96-01}, Theoretical Physics and
Particle Physics Experiments Divisions, G.~Altarelli, T.~Sj\"ostrand and
F.~Zwirner, eds., Vol. 2, pag. 3, Geneva, 19 February 1996.

\bibitem{cpcww} G.~Montagna, O.~Nicrosini and F.~Piccinini,
Comput. Phys. Commun. {\bf 90} (1995) 141.

\bibitem{wmass} Z.~Kunszt, W.J.~Stirling et al., ``Determination of the mass
of the $W$ boson'',
in {\it Physics at LEP2}, CERN {\bf 96-01}, Theoretical Physics and
Particle Physics Experiments Divisions, G.~Altarelli, T.~Sj\"ostrand and
F.~Zwirner, eds., Vol. 1, pag. 141, Geneva, 19 February 1996.

\bibitem{bhf} E.~N.~Argyres et al., Phys.~Lett.~{\bf B358} (1995) 339.

\bibitem{wwcd} W. Beenakker, F. Berends et al., ``WW cross-sections and
 distributions'',
in {\it Physics at LEP2}, CERN {\bf 96-01}, Theoretical Physics and
Particle Physics Experiments Divisions, G.~Altarelli, T.~Sj\"ostrand and
F.~Zwirner, eds.,  Vol. 1, pag. 79.

\bibitem{sf}{E.~A.~Kuraev and V.~S.~Fadin, Sov.~J.~Nucl.~Phys.
{\bf 41} (1985) 466;\\
G.~Altarelli and G.~Martinelli, in {\it Physics at LEP}, CERN Report
{\bf 86-02}, J.~Ellis and R.~Peccei eds. (CERN, Geneva, 1986); \\
O.~Nicrosini and L.~Trentadue, Phys.~Lett. {\bf B196} (1987) 551. }

\bibitem{jetset} T.~Sj\"ostrand, Comp. Phys. Commun. {\bf 82} (1994) 74; 
Lund University Report LU TP 95-20 (1995).

\bibitem{egdp} M.L.~Mangano, G.~Ridolfi et al., ``Event Generators for Discovery
Physics'', in {\it Physics at LEP2}, CERN {\bf 96-01}, Theoretical Physics and
Particle Physics Experiments Divisions, G.~Altarelli, T.~Sj\"ostrand and
F.~Zwirner, eds., Vol.2, pag.~299.

\bibitem{herwig} G.~Marchesini, B.~R.~Webber et al.,
Comput. Phys. Commun. {\bf 67} (1992) 465.

\end{thebibliography}
\end{document}